# Demonstration of mid-infrared waveguide photonic crystal cavities

Hongtao Lin,[1] Lan Li,[1] Fei Deng,[1] Chaoying Ni,[1] Sylvain Danto,[2] J. David Musgraves,[3] Kathleen Richardson,[2] Juejun Hu[1,*]

[1]*Department of Materials Science & Engineering, University of Delaware, Newark, DE 19716, USA*
[2]*College of Optics & Photonics, Department of Materials Science and Engineering,
University of Central Florida, Orlando, FL 32816, USA*
[3]*IRradiance Glass Inc. Orlando, FL 32828, USA*
*\*Corresponding author: [hujuejun@udel.edu](hujuejun@udel.edu)*

We have demonstrated what we believe to be the first waveguide photonic crystal cavity operating in the mid-infrared. The devices were fabricated from $Ge_{23}Sb_7S_{70}$ chalcogenide glass on $CaF_2$ substrates by combing photolithographic patterning and focus ion beam milling. The waveguide-coupled cavities were characterized using a fiber end fire coupling method at 5.2 μm wavelength, and a loaded quality factor of ~ 2,000 was measured near the critical coupling regime.

Waveguide nanobeam photonic crystal (PhC) cavities have become a promising alternative to conventional cavity geometries [1, 2]. Quality factors (Q) up to $10^6$ have been achieved in these 1D cavities through a Bloch mode engineering design approach [3]. Coupled with their small mode volume, PhC nanobeam cavities are recognized as an ideal platform for exploring cavity-enhanced photon-matter interactions given their high cavity finesse [4]. In addition, PhC nanobeam cavities are inherently amenable to integration with traditional index-guided waveguides, an important advantage for planar photonic integration. To date, waveguide PhC cavities have only been characterized at visible or near-infrared (near 1550 nm) wavelengths [5-9]. Waveguide PhC cavity devices operating in the mid-infrared (mid-IR, 3 to 20 μm wavelengths), a strategically important wave band for spectroscopic sensing, free space communications, and thermal imaging [10], have not yet been demonstrated.

A main technical challenge to mid-IR photonic device fabrication is the much limited material choices. The conventional optical cladding material, silica, becomes opaque at wavelengths longer than 3.5 μm. As a consequence, mid-IR devices demonstrated to date almost exclusively rely on wafer bonding to mid-IR transparent substrates (e.g. sapphire [11-13] or silicon nitride [14]) or suspended structures [15-19], which significantly complicate device fabrication and integration. Here we explore an alternative device design based on amorphous chalcogenide glasses (ChGs). These glass materials are well known for their broad transparency in the mid-IR range [20]. Furthermore, their amorphous structure enables direct monolithic integration on virtually any substrates free of lattice matching constraints. Finally, ChGs possess a photothermal figure-of-merit 100 times higher than those of silica and silicon, making them ideal material candidates for ultra-sensitive photothermal spectroscopic sensing applications [21, 22], where mid-IR waveguide PhC cavities constitute the basic device building block. Recently, we have demonstrated chalcogenide glass resonators on silicon with a high intrinsic Q-factor of $2 \times 10^5$ in the mid-IR following the monolithic integration approach [23].

In this letter, we discuss the first experimental demonstration of mid-IR 1D waveguide photonic crystal cavities. The cavities were made of $Ge_{23}Sb_7S_{70}$ chalcogenide glass, and were monolithically fabricated on mid-IR transparent $CaF_2$ substrates. The $Ge_{23}Sb_7S_{70}$ glass is transparent up to 12 μm wavelength, and is stable against moisture, oxidation, and acid solutions. The combination of high-index $Ge_{23}Sb_7S_{70}$ glass (n = 2.2) waveguide core and low refractive index $CaF_2$ substrate (n = 1.4) offers large index contrast for strong optical

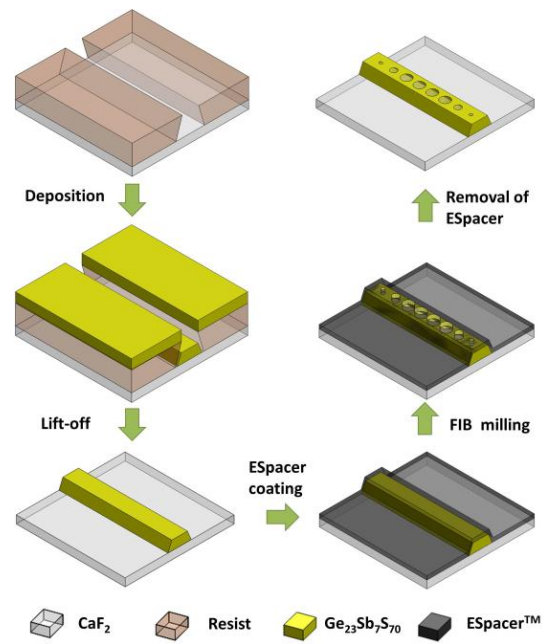

Fig. 1. Schematic fabrication process flow for the $Ge_{23}Sb_7S_{70}$ chalcogenide glass waveguide photonic crystal cavities on $CaF_2$.

confinement in the cavities.

Figure 1 schematically illustrates the fabrication process flow of the 1D waveguide photonic crystal cavity. The $Ge_{23}Sb_7S_{70}$ glass films were first deposited on 1" (111) $CaF_2$ substrates (Shanghai Daheng Optics and Fine Mechanics) via thermal evaporation. The glass films show excellent adhesion with the $CaF_2$ substrates, and no film delamination was observed during subsequent processing. The PhC nanobeam was fabricated using a two-step process combining UV-lithography and focused ion beam (FIB) milling. Instead of sculpting the entire structure using FIB, the two-step process minimizes milling area and improves fabrication throughput. Details of the glass film deposition and lithographic patterning processes can be found elsewhere [24, 25]. After fabrication of 3 µm width, 1.2 µm thick single mode waveguide devices, a layer of 20 nm thick water-soluble conducting polymer (Espacer™, Showa Denko) was spin coated onto the substrate to prevent charging [26, 27] during ion beam milling. The PhC holes were defined by a $Ga^{2+}$ ion beam (beam current 20 pA, accelerating voltage 30 kV) using a Zeiss Auriga 60 CrossBeam™ FIB nanoprototyping workstation. After milling, the PhCs were rinsed in deionized water to remove the Espacer layer and complete the device fabrication.

Fig. 2a gives an SEM anatomy view of holes etched by FIB. From left to right, the milling doses were linearly increased from 0.2 nC/µm² to 1.2 nC/µm². The etch depths linearly increase with the etch dose (Fig. 2c) and the effective etch yield is about 1.2 µm³/nC. We also note that $CaF_2$ serves as an excellent etch stop to $Ga^{2+}$ ion milling, which allows precise definition of the PhC hole depth. A close view in Fig. 2b clearly shows that there is little re-deposition of glass on the hole sidewalls, yielding a high-quality, smooth surface finish.

The fabricated cavity geometry consists of a segment of glass waveguide sandwiched between two identical PhC mirror reflectors milled at 0.4 nC/µm² ion beam dose. Figure 2d presents an optical microscope top-view image of the cavity structure, and Fig. 2e shows an SEM top view of one of the PhC mirrors. We chose Fabry–Pérot cavity designs with a long cavity length in this initial demonstration to facilitate isolation and quantification of optical loss mechanisms in the cavity, since the mirror strength (reflectance) and extrinsic cavity Q-factor can be readily tuned by changing the number of holes comprising the mirrors. To minimize optical scattering loss, 5 size-tapered holes are inserted at both sides of the PhC mirror. The diameter of PhC mirror holes was fixed at 1.2 µm, and the diameters of the 5 taper holes were linearly decreased from 1 µm to 0.2 µm, as illustrated in Fig. 2f. Our numerical simulations indicate a reduction of single-bounce scattering loss from 4% to 0.5% though incorporation of the taper structure.

We fabricated two types of cavities with cavity lengths of 170 µm and 440 µm, respectively. The hole period for both types of cavities was 1.65 µm, which creates a photonic band edge near 5.2 µm wavelength. Strong wavelength dependence of PhC mirror reflectance is thus expected in the wavelength regime near the band edge, leading to significant reduction of the cavity Q-factor as the resonant wavelength approaches the band edge. Fig. 3 plots transmission and reflection spectra of a waveguide PhC mirror with a period of 1.65 µm simulated using the Finite-Difference Time-Domain (FDTD) technique.[28] It can be seen that the PhC mirror strength decreases at longer wavelength, manifesting the PhC band edge effect.

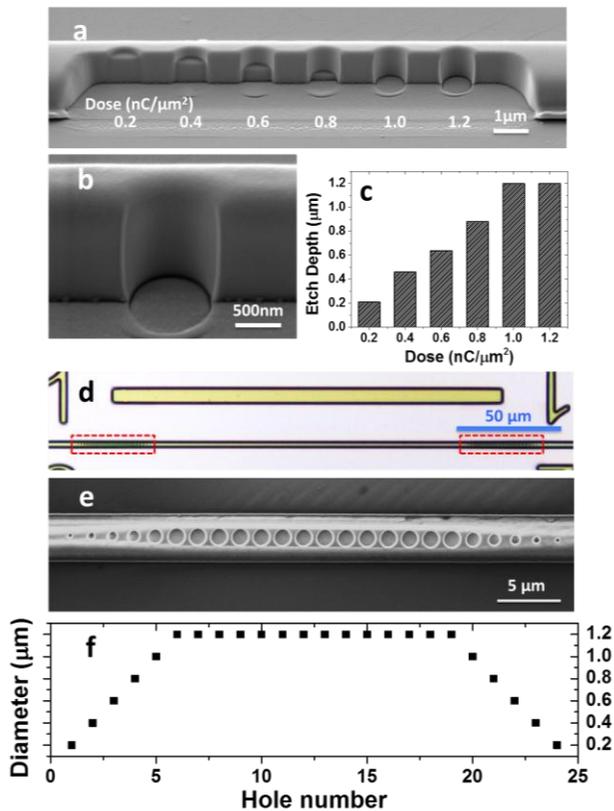

Fig. 2. (a) An anatomy section of holes in a 3 µm width, 1.2 µm thick $Ge_{23}Sb_7S_{70}$ chalcogenide glass waveguide milled using different ion beam dose for etch rate calibration; (b) cross-sectional SEM image of a photonic crystal through hole, showing a smooth surface finish and near vertical sidewalls; (c) FIB etch depth of the holes plotted as a function of ion beam dose inferred from Fig. 2a; the $CaF_2$ substrate serves as an excellent etch stop for FIB milling and thus the etch depth saturates at the glass film thickness; (d) top view microscope image of waveguide photonic crystal cavity, consisting of a section of unstructured channel waveguide (~ 170 µm in length) confined between two PhC mirrors (marked by the red boxes); (e) top-view SEM image of one of the photonic crystal mirrors; (f) diameters of photonic crystal mirror holes shown in Fig. 2e, the center-to-center spacing between the holes is fixed at 1.65 um.

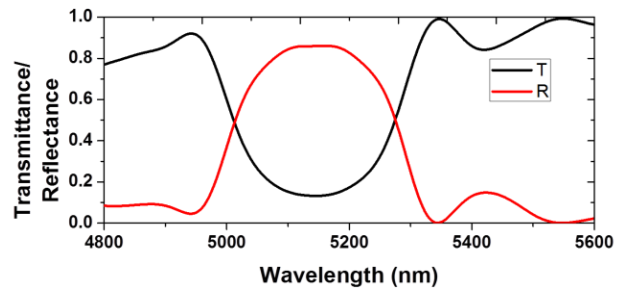

Fig. 3. Simulated transmission and reflection spectra of the photonic crystal mirror in Fig. 2e.

On the other hand, in the spectral regime near the center of the photonic band gap, little wavelength-dependence of the cavity Q-factor is expected, and the mirror strength can be effectively tuned by changing the number of PhC holes: increasing the hole number leads to high mirror reflectance and an increase of the extrinsic cavity Q-factor.

These general characteristics of the PhC cavities were validated through transmission measurements of the PhC cavities. The measurements were performed on a home-built fiber end fire coupling system. Details of the experimental set up and measurement method were discussed in our previous work [23]. Figure 4a shows the transmission spectrum of a waveguide PhC cavity with a cavity length of 170 μm. Multiple longitudinal orders of cavity resonances were clearly visible from the spectrum. The loaded cavity Q-factors of these resonances monotonically decrease from 1,300 to 400 as the resonant wavelength approaches the photonic band edge (Fig. 4b), which agrees well with our simulation results and unequivocally confirms the photonic band gap effect in our fabricated structures.

To further assess the loss mechanisms in the waveguide photonic crystal cavities, a series of cavities with different PhC mirror hole numbers were fabricated and tested. As we expected, the Q factors show minimal wavelength dependence, the level of the dependence being completely overshadowed by our experimental measurement uncertainty. Figure 5a plots the transmission spectrum of a photonic crystal cavity with 10-hole mirror sets. Figure 5b compares the experimentally measured and numerically simulated quality factors as the PhC mirror hole number is varied. The measured Q-factors were averaged from at least 7 resonance peaks near the center of the photonic stop band. The figure shows monotonic increase of loaded cavity Q-factor as the mirror hole number increases, which is an anticipated result since external Q-factor scales with cavity mirror strength. The cavity parameters listed in Table 1 were fitted from the measured spectrum and FDTD simulation results using the following equation:

$$T = \left| \frac{T_s \, exp(i \, k \, L - 0.5 \alpha L)}{1 - R_s \, exp(2 \, i \, k \, L - \alpha L)} \right|^2$$

where T is the total transmittance through the cavity, $T_s$ and $R_s$ are the mirror transmittance and reflectance values derived from FDTD simulations, $\alpha$ represents the optical loss which includes both waveguide propagation loss and scattering loss of the mirrors, L gives the cavity length, and k denotes the wave vector and is defined by $k = 2 \pi \, n_{eff} / \lambda$. The wavelength-dependent waveguide effective index $n_{eff}$ can be obtained from modal simulations. The high optical loss ($\alpha$ = 50 dB/cm), which accounts for the relatively low cavity Q, may be attributed to scattering loss resulting from waveguide sidewall roughness and the PhC mirrors. Further improvement of the waveguide photonic crystal cavity Q-factors is expected through processing optimization and new designs following a deterministic nanobeam cavity optimization strategy [29].

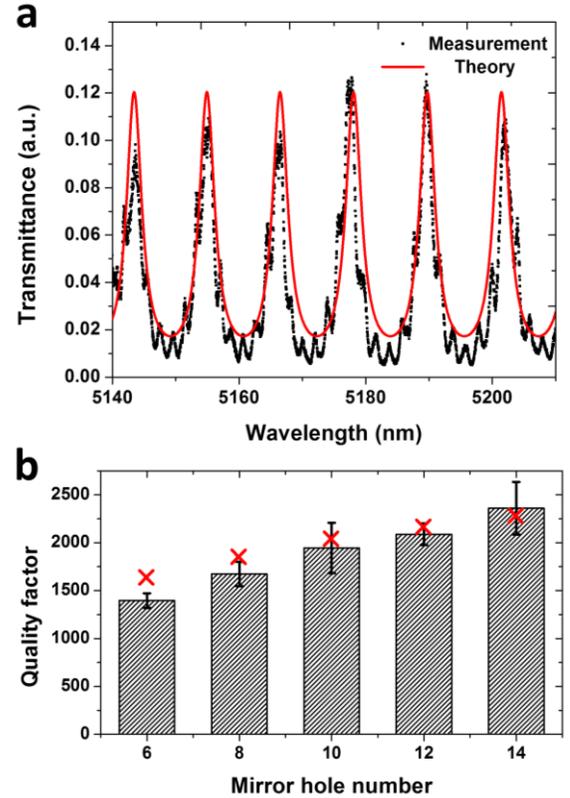

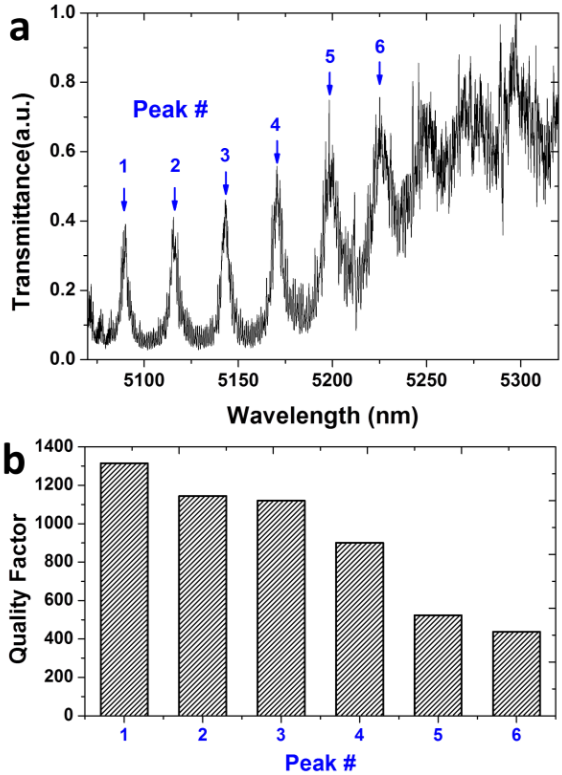

Fig. 4. (a) Mid-IR optical transmission spectrum of the waveguide photonic crystal cavity shown in Fig. 2d measured using a wavelength sweeping method; (b) the quality factor of resonance peaks in Fig. 4a; their Q-factors monotonically decreases as their resonant wavelengths approach the photonic band edge.

Fig. 5. (a) Mid-IR optical transmission spectrum of a waveguide photonic crystal cavity with a 440 μm-long cavity positioned between two 10-hole PhC mirrors; (b) the evolution of loaded cavity Q-factors as the PhC mirror hole number increases: the black bars represent experimentally measured Q values, and the red crosses denote FDTD simulation results.

Table. 1. Parameters used to calculate the PhC cavity transmittance in Fig. 5a

| $T_S$ | $R_S$ | $n_{eff}$ |
|---|---|---|
| 0.2463 | 0.7489 | 1.6524 |
| $n_g$ | $\alpha$ | $L$ (μm) |
| 2.6083 | 50 dB/cm | 442 |

In conclusion, we have shown, to the best of our knowledge, the first demonstration of mid-IR waveguide photonic crystal cavities. The cavity devices are made of $Ge_{23}Sb_7S_{70}$ chalcogenide glass on $CaF_2$ substrate using a two-step patterning process combining lithography and focused ion beam milling techniques. The devices exhibit a loaded quality factor of ~ 2,000, and an extinction ratio up to 13 dB near the critical coupling operation regime at the mid-IR wavelength of 5.2 μm. The PhC cavity device platform can potentially serve as a useful building block for applications including on chip chemical sensing, optical free-space communications, and thermal imaging.

**Acknowledgement:** The authors gratefully acknowledge support from the National Science Foundation under award number 1200406 and EPSCoR Grant number EPS-0814251. Additional partial support has been provided by US Department of Energy [Contract # DE-NA000421], NNSA/DNN R&D.